\documentclass[aps,pra,groupedaddress,notitlepage]{revtex4-1}

\usepackage{amssymb}
\usepackage{amsmath}
\usepackage{amsfonts}
\usepackage{amsthm}
\usepackage{graphicx}
\usepackage{color}
\usepackage{bm}
\usepackage{url}
\usepackage{hyperref}

\begin{document}

\title{Analog-Based Forecasting of Turbulent Velocity:\\ Relationship between Unpredictability and Intermittency}

\author{E. Frog\'e$^{1,2}$}
\email{Correspondence: ewen.froge@imt-atlantique.fr}

\author{C. Granero-Belinch\'on$^{1,2}$} 
\author{S. G. Roux$^{3}$}
\author{N. B. Garnier$^{3}$}
\author{T. Chonavel$^{1}$}

\affiliation{%
$^{1}$ Department of Mathematical and Electrical Engineering, IMT Atlantique, Lab-STICC, UMR CNRS 6285, 655 Av. du Technop\^ole, Plouzan\'e, 29280, Bretagne, France.}
\affiliation{$^{2}$ Odyssey, Inria/IMT Atlantique, 263 Av. G\'en\'eral Leclerc, Rennes, 35042, Bretagne, France.}
\affiliation{$^{3}$ Univ Lyon, ENS de Lyon, CNRS, Laboratoire de Physique, 46, all\'ee d'Italie, Lyon, F-69364, Rh\^one-Alpes, France}

\begin{abstract}
This study evaluates the performance of analog-based methodologies to predict, in a statistical way, the longitudinal velocity in a turbulent flow. The data used comes from hot wire experimental measurements from the Modane wind tunnel. We compared different methods and explored the impact of varying the number of analogs and their sizes on prediction accuracy. We illustrate that the innovation, defined as the difference between the true velocity value and the prediction value,  highlights particularly unpredictable events that we directly link with extreme events of the velocity gradients and so to intermittency. A statistical study of the innovation indicates that while the estimator effectively seizes linear correlations, it fails to fully capture higher-order dependencies. The innovation underscores the presence of intermittency, revealing the limitations of current predictive models and suggesting directions for future improvements in turbulence forecasting.
\end{abstract}

\maketitle 

\section{Introduction}

For a given temporal stochastic process $X$, Cram\'er defined {\em innovation} to describe the new random inputs into the system and to quantify the unpredictability in the time evolution of $X$~ \cite{cramerLinearPredictionProblem1960,doobStochasticProcesses1953}. 
For a deterministic dynamical system, if a solution exists and is known, the lack of noise leads to a vanishing innovation. However, in chaotic systems, innovation rarely vanishes and can be used to quantify the degree of predictability of the process and to identify unpredictable localized events related to the system's dynamics.

Fluid turbulence is one example of a chaotic system with very complex multiscale dynamics, in which the velocity field is impossible to derive deterministically \cite{peinkeChaosFractalsTurbulence1993, crisantiIntermittencyPredictabilityTurbulence1993, crisantiPredictabilityVelocityTemperature1993, aurellPredictabilitySystemsMany1996, boffettaPredictabilityChaoticSystems1998, boffettaChaosPredictabilityHomogeneousIsotropic2017}. Turbulent velocity presents long-range correlations, indicating non-local interactions~\cite{frischTurbulenceLegacyKolmogorov1995b}, and intermittency leads to extreme events in the velocity gradients~\cite{frischTurbulenceLegacyKolmogorov1995b,meneveauMultifractalNatureTurbulent1991}. The chaotic dynamics of turbulence showcases extreme sensitivity to initial conditions, where a small noise due to experimental errors or numerical approximations, rapidly amplify \cite{lorenzAtmosphericPredictabilityRevealed1969a}. Small thermal fluctuations can also lead to different solutions \cite{ruelleMicroscopicFluctuationsTurbulence1979}. Unpredictability in turbulent flows is an inherent feature tied to the stochasticity of their dynamics~\cite{biferaleOptimalSubgridScheme2017,bandakSpontaneousStochasticityAmplifies2024}. As a consequence, predictability in turbulence should be addressed with probabilistic forecasting methods that produce possible future states modeled by a probability density function (PDF)~\cite{liStochasticReconstructionGappy2024}.

We propose to characterize the predictability of turbulent velocity in terms of innovation, highlighting that difficult-to-predict events are related to bursts in velocity gradients and thus to intermittency. In this context, we adapt analog-based estimators\cite{lorenzAtmosphericPredictabilityRevealed1969a} previously used in meteorological \cite{doolNewLookWeather1989a,tothLongRangeWeatherForecasting1989} and dynamical systems applications~\cite{sugiharaNonlinearForecastingWay1990} to predict the next value in a hot-wire velocity measurement from a grid turbulence experiment in the Modane wind tunnel~\cite{kahalerrasIntermittencyReynoldsNumber1998a}. 
Very importantly the sensitivity of turbulent dynamics to very small disturbances means that close initial conditions may still yield widely different trajectories. This highlights the importance of probabilistic forecasting approaches. Consequently, we use analog-based approaches to estimate the mean and variance of a PDF characterizing the next turbulent velocity, that define respectively the predicted velocity and its associated uncertainty.
We compute the corresponding innovation as the difference between the real signal and the mean analog prediction. We then analyze the statistical properties of the velocity, velocity increment, innovation, and the cumulative sum of innovations by examining their second-order structure function and flatness. On the one hand, the second order structure function is a second-order statistic that characterizes the energy distribution across scales. On the other hand, the flatness is a higher-order statistic that characterizes the significance of extreme events across scales.

While the analog-based innovation predictions can effectively capture and predict non-extreme dynamics by removing some linear dependencies, extreme events in velocity gradients due to intermittency are not accurately predicted by the analog method. This method falls short in accounting for higher-order dependencies. The innovation process shows a nearly white spectrum with a non-Gaussian, fat-tailed probability distribution and retains higher-order dependencies. Extreme events in innovation are localized in time and appear concomitantly with extreme velocity gradient events. Moreover, these extreme events are also concomitant with higher values of the variance of the analog forecast, and so of the variance of the innovation. Intermittency leads to unpredictability of turbulent velocity in terms of both the mean prediction and its associated uncertainty.

This article is structured into three sections. In section \textit{Turbulence \& Intermittency} we discuss intermittency, its characterization, and the turbulent data used. Section \textit{Innovation} introduces the new proposed framework and the estimators we compute. Finally, section \textit{Results} evaluates the performance of the different estimators and discusses the statistics of innovation on real experimental data.

\section{Turbulence \& Intermittency}
\label{sec:Intermittency}

The Richardson cascade picture of three-dimensional turbulence identifies three domains of scale: the integral domain, which contains the large scales where energy is injected; the inertial range, where energy cascades from large to small scales; and the dissipative domain where energy is dissipated at smaller scales.

In the case of three-dimensional fully developed turbulence, the multifractal formalism~\cite{frischSingularityStructureFully1985,frischTurbulenceLegacyKolmogorov1995b,paladinAnomalousScalingLaws1987} describes in the inertial domain a power law behavior of the structure functions $S_p(l)$ of the turbulent velocity field as a function of the scale:
\begin{equation}
\label{eq:Structure}
    S_p(l) = \mathbb{E}[(\delta_l v)^p] \propto l^{\zeta(p)}
\end{equation}

\noindent where $\delta_l v(x) = v(x+l)-v(x)$ is the velocity increment of size $l$ and $\zeta(p)$ is the scaling exponent.

The Kolmogorov's 1941 (K41) theory~\cite{kolmogorovLocalStructureTurbulence1991} originally proposed a linear scaling exponent $\zeta(p) = p h$, with a single Holder exponent $h = \frac{1}{3}$ i.e. a monofractal velocity field. This model does not take into account the intermittency phenomenon highlighted by experiments~\cite{gagneNewUniversalScaling1990,kahalerrasIntermittencyReynoldsNumber1998a}. Intermittency is characterized by a deformation of the probability density function of the velocity increments from Gaussian at large scales to non-Gaussian with heavy tails at small scales~\cite{castaingVelocityProbabilityDensity1990}. This behavior reflects the presence of extreme events, which become more significant at smaller scales. It implies multifractality: a non-linear scaling exponent $\zeta(p)$ with a dependancy of h upon p $\zeta(p)=ph(p)$. An intermittent model of turbulence was proposed by Kolmogorov and Obukhov (KO62) ~\cite{kolmogorovRefinementPreviousHypotheses1962, obukhovSpecificFeaturesAtmospheric1962} and later studied within the multifractal formalism~\cite{chevillardPhenomenologicalTheoryEulerian2012,castaingVelocityProbabilityDensity1990,paladinAnomalousScalingLaws1987,castaingLogsimilarityTurbulentFlows1993,delourIntermittency1DVelocity2001a,meneveauMultifractalNatureTurbulent1991}.

The energy distribution across scales is characterized by the second-order structure function $S_2(l)$ which, for scales in the inertial domain, is proportional to $l^{2/3}$ up to intermittent corrections~\cite{frischTurbulenceLegacyKolmogorov1995b}. $S_2(l)$ characterizes second-order statistics of the turbulent velocity field in the same way as autocorrelation and power spectrum~\cite{frischTurbulenceLegacyKolmogorov1995b}.

The flatness characterizes the relative importance of extreme events and allows for a measure of the deformation of the PDF. It is defined for a given scale $l$ as  \\

\begin{equation}
F(l) = \frac{S_4(l)}{3S_2(l)^2}
\end{equation}

It thus characterizes higher-order statistics of velocity. In a monofractal Gaussian field, the flatness  remains constant across scales and is equal to $1$. In contrast, turbulence shows an increasing flatness with decreasing scale $l$, indicating a higher prevalence of extreme events or bursts at smaller scales~\cite{dubrulleKolmogorovCascades2019,buariaExtremeVelocityGradients2019,moffattExtremeEventsTurbulent2021}. The increased flatness at smaller scales due to the departure from Gaussian statistics is a key indicator of intermittency. 

\subsection{Modane Turbulent Velocity Dataset}

We use an Eulerian longitudinal velocity measurement from an experimental grid turbulence setup in the Modane wind tunnel~\cite{kahalerrasIntermittencyReynoldsNumber1998a}. The full velocity measure spans $1000$ seconds with a sampling frequency of $f_s=25$ kHz. Velocity measurements were obtained using hot-wire anemometry.
The Taylor scale Reynolds number of the flow is $R_\lambda = 2500$.

To relate temporal measurements to spatial properties, we employ Taylor's hypothesis of frozen turbulence, which allows us to interpret temporal variations in the velocity signal as spatial variations using 
$v(x, t) = v(x - Vt, 0)$
, where $v$ is the velocity, $x$ is the space coordinate in the longitudinal direction, defined along the mean velocity of the flow, $t$ is the time coordinate and $V=20.5 m s^{-1}$ is the mean velocity of the flow. 

From previous studies, the integral scale of the flow is $L = 2350 dl$ and the Kolmogorov scale \cite{kolmogorovLocalStructureTurbulence1941a} of the flow is $\eta = 5 dl$~\cite{granero-belinchonScalingInformationTurbulence2016}, where $dl=V/f_s $.
In the following, all results are presented in function of $t/T$ with $T=L/V$.

\section{Innovation}
\label{sec:method}

\subsection{Theoretical Description}

For a stochastic process $X=\{x_t\}$ sampled at intervals equally separated by $dt$, we define the innovation $\epsilon_{t}$ as: 

\begin{equation}
    \epsilon_{t}=x_t-\mathbb{E}[x_{t}|x_{-\infty:t-dt}]
\label{eq:Innovation}
\end{equation}

where $\mathbb{E}[x_{t}|x_{-\infty:t-dt}]$ is the expected value of $x_t$ conditioned on its complete past $x_{-\infty:t-dt}$. 

We interpret $\epsilon_{t}$ as the novel, unpredictable component of $x_t$ at time $t$~\cite{cramerLinearPredictionProblem1960,doobStochasticProcesses1953}. Thus, innovation acts as a limit to the error of an optimal prediction, providing insight into the degree of unpredictability of the system as a function of time. 
Very interestingly, innovation can point out specific unpredictable events localized in time.

In practice, the expected value $\mathbb{E}[x_{t}|x_{-\infty:t-dt}]$ cannot be conditioned on an infinite past and we restrict the conditioning  on a finite past interval $t\in[t-p dt, t-dt]$, leading to the approximation: 

\begin{equation}
    \epsilon_{t}^{(p)}=x_t-\mathbb{E}[x_{t}|x_{t-p dt:t-dt}]
\label{eq:Innovation_p}
\end{equation}

\noindent where $p$ indicates the finite duration of the past horizon.

\subsection{Analogs}

For a given time $t$, we consider the temporal sub-sequence $\vec{x}^{(p)}_{t}$ of size $p$ of the stochastic process $X$ as:

\begin{equation}
    \vec{x}_{t}^{(p)} = 
    \begin{pmatrix}
        x_{t-dt} \\
        \vdots \\
        x_{t-p dt}
    \end{pmatrix}
\end{equation}

Two such sub-sequences $\vec{x}_{t}^{(p)}$ and $\vec{x}_{t'}^{(p)}$ considered at two different times $t$ and $t'$ are called analogs if they are close enough in terms of some given $p$-dimensional distance. 

Based on Poincar\'e's theorem~\cite{poincareProblemeTroisCorps1890a}, we assume that 1) analogs exist if the time series is sufficiently long; and 2) close analogs will lead to close successors. Then for a given $t$, the set of successors $x_{t'}$ of the analogs $\vec{x}_{t'}^{(p)}$ of $\vec{x}_{t}^{(p)}$ can be used to predict in a statistical way the successor $x_t$. This approach relies on the similarity of past sequences and their subsequent outcomes for making predictions.

In the context of the present study, analog forecasting serves as a methodological framework for estimating the probability density function $q(x_{t}|x_{t-p dt:t-dt})$ and its associated expected value $\mathbb{E}[x_{t}|x_{t-p dt:t-dt}]$ that appears in the definition of innovation.

\subsection{Analog-based predictions and innovations }
\label{subsec:Estimator}

Given $\vec{x}_{t}^{(p)}$ and a fixed integer $k$, we define the analogs of $\vec{x}_{t}^{(p)}$ as the set $\{ \vec{x}_{t'_i}^{(p)} \}_{1\le i\le k}$ of the $k$ historical sequences with $t'_i<t$ that are the closest to $\vec{x}_{t}^{(p)}$ in terms of the Euclidean distance, as suggested in~\cite{platzerUsingLocalDynamics2021c,lguensatAnalogDataAssimilation2017a}. 


The mean $\hat{x}_t^{(p)}$ and variance $\sigma_t^2$ of the analog-based estimation of $q(x_{t}|x_{t-p dt:t-dt})$ are derived from the set of selected analogs and represent respectively the prediction $\hat{x}_t^{(p)}=\mathbb{E}[x_{t}|x_{t-p dt:t-dt}]$ and its uncertainty. Different estimators for $\hat{x}_t^{(p)}$ and $\sigma^2_t$ exist in the literature \cite{platzerUsingLocalDynamics2021c}.


\subsection{Average}
The original method for analog prediction computes a weighted average of the successors $\{ x_{t'_i}\}_{1 \le i \le k}$ of the analogs ~\cite{kruizingaUseAnalogueProcedure1983,monacheKalmanFilterAnalog2011}.
This leads to the prediction $\hat{x_t}^{(p)}$ with variance $\sigma^2_t$:
\begin{equation}
\label{eq:weighted_average}
     \hat{x_t}^{(p)} = \sum_{i=1}^{k} w_{ii} x_{t'_i},\quad \sigma^2_{t}= \sum_{i=1}^{k} w_{ii} \left( x_{t'_i}-\hat{x_t}^{(p)} \right)^2
\end{equation}
with the corresponding innovation $\hat{\epsilon}_t^{(p)}=x_t -\hat{x_t}$.

The weights $w_{ii}$ are defined as the components of a diagonal weight matrix $\bm{W}$:
\begin{equation}\label{eq:weights}
    w_{ii} = \frac{e^{-||\vec{x}_{t}^{(p)} - \vec{x}_{t'_i}^{(p)}||/\lambda}}{\sum_{j=1}^{k} e^{-||\vec{x}_{t}^{(p)} - \vec{x}_{t'_j}^{(p)}||/\lambda}}
\end{equation}

\noindent with $\lambda=\text{median}(\{||\vec{x}_{t}^{(p)}-\vec{x}_{t'_i}^{(p)}||\}_{1 \le i \le k})$, following~\cite{platzerUsingLocalDynamics2021c,lguensatAnalogDataAssimilation2017a}. 

This averaging method tends to draw predictions towards the average of the $k$ successors,  resulting in estimations of the innovation that are higher in unexplored regions\cite{lguensatAnalogDataAssimilation2017a}. 

\subsection{Linear Regression (LR)}
Following~\cite{platzerUsingLocalDynamics2021c}, we then make an analog-based prediction using a weighted linear regression where the analogs are used as explanatory variable and their successors are the explained one. 

For a given number $k$ of analogs, the minimization problem can be written as $\min\limits_{\beta_t}||\vec{y}-\bm{W}\bm{X_{p}}\beta_t||$ where:

\begin{itemize}

\item $\vec{y}$ is the vector of analog successors and $\bm{X_p}$ is a $k \times (p+1)$ matrix containing all analog vectors with a first column of ones: 

\begin{equation}
    \label{eq:matrix_definition}
    \vec{y}=\begin{pmatrix}
        x_{t_1'}\\
        \vdots \\
        x_{t_k'} 
    \end{pmatrix}, \, \bm{X_p}=
     \begin{pmatrix}
        1 & x_{t'_1-dt} & \hdots & x_{t'_1-pdt} \\
        \vdots & \vdots & & \vdots \\
        1 & x_{t'_k-dt} & \hdots & x_{t'_k-pdt}
    \end{pmatrix} 
\end{equation}

\item $\bm{W}$ is a $k \times k$ diagonal matrix which specifies the weight of each analog. The non-zero elements $w_{ii}$ of $\bm{W}$ are defined using eq.\ref{eq:weights}.

\item $\beta_t$ is the coefficients vector of the weighted linear regression that depend on $t$ through the definition of the analogs, that we search for.

\end{itemize}

The ordinary least squares estimator for the minimization gives:
\begin{equation}
\label{eq:inversion}
    \hat{\beta_t}=(\bm{X_p}^T \bm{W} \bm{X_p})^{-1} \bm{X_p}^T \bm{W} \vec{y}
\end{equation}

The analog prediction $\hat{x_t}^{(p)}$ with variance $\sigma^2_{t}$ then reads:
\begin{equation}
\label{eq:innovation}
\hat{x_t}^{(p)}\equiv\hat\beta_t^T \vec{x}_{t}^{(p)},\quad \sigma^2_{t} = \sum_{i=1}^{k} w_{ii} \left( \xi_{t,i} \right)^2
\end{equation}
where $\vec{\xi}_t= \vec{y}-\bm{W}\bm{X_p}\hat\beta_t$ are the residues, and the corresponding innovation is $\hat{\epsilon}_t^{(p)}=x_t-\hat\beta_t^T \vec{x}_{t}^{(p)}$.

This linear regression approach in phase space establishes a relationship from past states to their successors based on historical analogs. It has two parameters: the number $k$ of neighbors and the dimension $p$ of the sub-sequences. Note that for the inversion (eq.\ref{eq:inversion}) to be possible, one should have $k\ge p$.

\subsection{Normalized LR}
One can focus on trends rather than values by removing the mean value of the sub-sequences, allowing the analog search to prioritize patterns over analogs' values~\cite{barnettMultifieldAnalogPrediction1978}. This method is thus identical to the linear regression method after replacing all coordinates $x_{t_j}$ of sub-sequences $\vec{x}_{t}^{(p)}$ by $x_{t_j}- \overline{\vec{x}_{t}^{(p)}}$ with $ \overline{\vec{x}_{t}^{(p)} }\equiv\frac{1}{p}\sum_{1\le j \le p}x_{t - jdt}$.

\section{Results}
\label{sec:results}
\subsection{Analogs Prediction}

In this section, we compare the performance of the three analog-based predictions presented above when used on Modane turbulent velocity signal $v$. We also evaluate the impact of $k$ and $p$ in the performance, in order to determine the optimal values for $k$ and $p$.

\begin{figure}[t]
    \centering
    \includegraphics[width=0.6\linewidth]{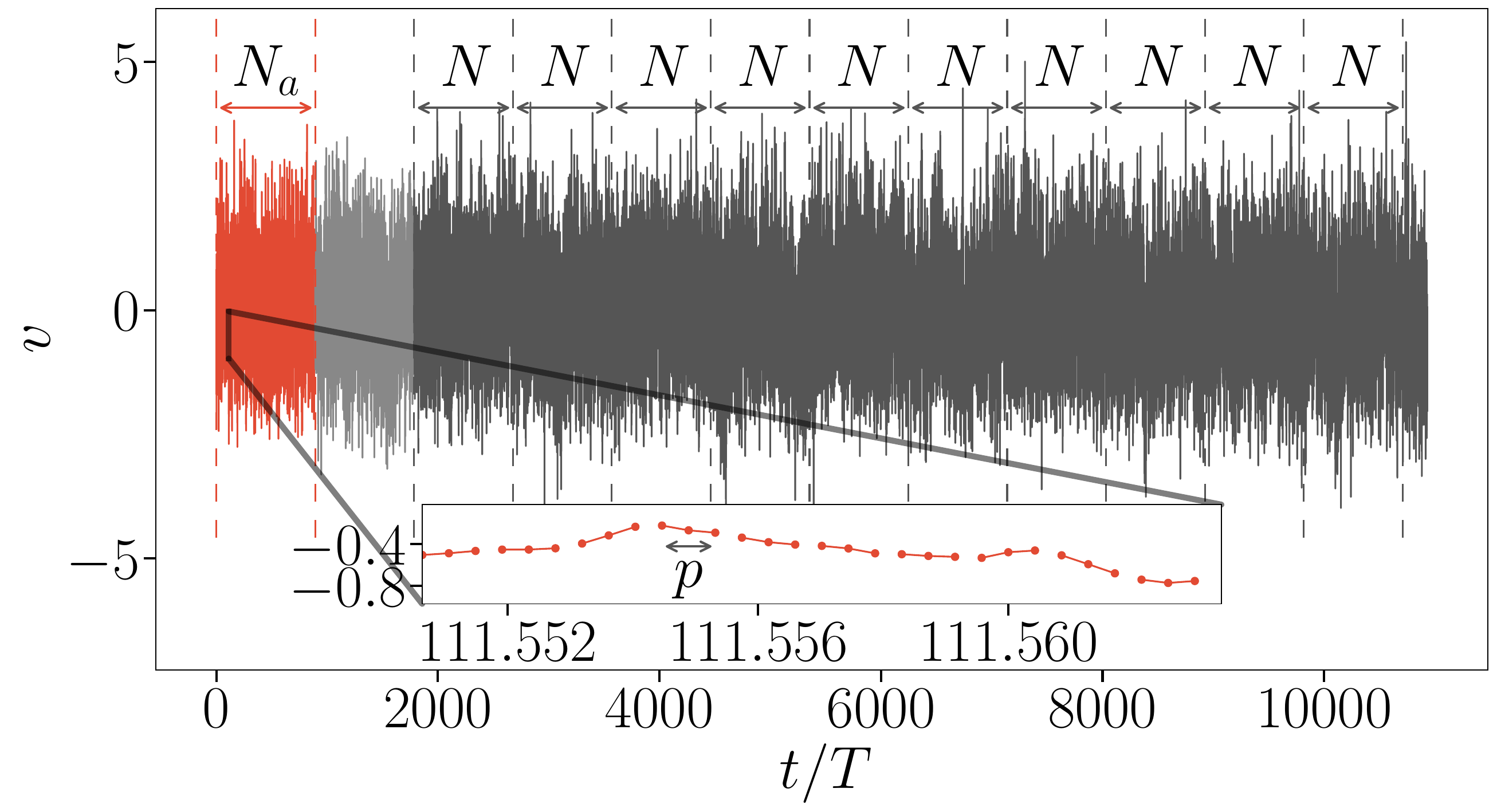}
    \caption{Visualization of Modane turbulent velocity data usage for the innovation $\epsilon_t^{(p)}$. The first red section of size $N_{\rm a}$ represents the historical record used for analog search. The subsequent ten black sections are the $n_{\rm real}=10$ realizations of size $N$ over which the innovation is computed. Vertical dashed lines indicate the boundaries of the historical record and of the realizations. The inset shows a zoom of the historical record with non-overlapping analogs candidates and a highlighted analog of length $p=3$.}  

    \label{fig:dataset_configuration}
\end{figure}

We compute the corresponding innovations $\epsilon_t^{(p)}$  for different methodologies and $k$ and $p$ combinations. This is done on $n_{\rm real}= 10$ realizations of length $N=2^{21}\simeq 900 T$ samples of $v$, see figure \ref{fig:dataset_configuration}. The database containing the historical record for analog searching is chosen to precede and not overlap with the portion over which the predictions and innovations are computed and it contains a total of $N_{\rm a}=N$ samples, where $N_{\rm a}$ has been selected to ensure that further increasing  the size of the historical record does not significantly affect the results. Overlap between analogs is not permitted, ensuring that the selected analogs are distinct. The results are computed over the $n_{\rm real}$ realizations.

In order to quantify the performance of the estimator over a full time series, we compute the Mean Squared Error (MSE) between the Modane turbulent velocity and the mean prediction $\frac{1}{N}\sum_{j=1}^{N}(v_j-\hat{v_j}^{(p)})^2$. This MSE is equal to the variance in time of the innovation $\epsilon_t$. 

We can consider the naive prediction $\hat{v_t} = v_{t-dt}$ which assumes a locally constant speed and hence doesn't account for any dynamics or even any change in the process. This assumption leads to the MSE: $\frac{1}{N}\sum_{j=1}^{N}(v_j-v_{j-1})^2$, which is nothing but the variance of the increment $\delta_{dt} v$. As a consequence, we can use the increment as a baseline of the behavior of the innovation $\epsilon_t^{(p)}$ for the three analog-prediction methods presented earlier.

\begin{figure}[t]
    \centering
    \includegraphics[width=0.6\linewidth]{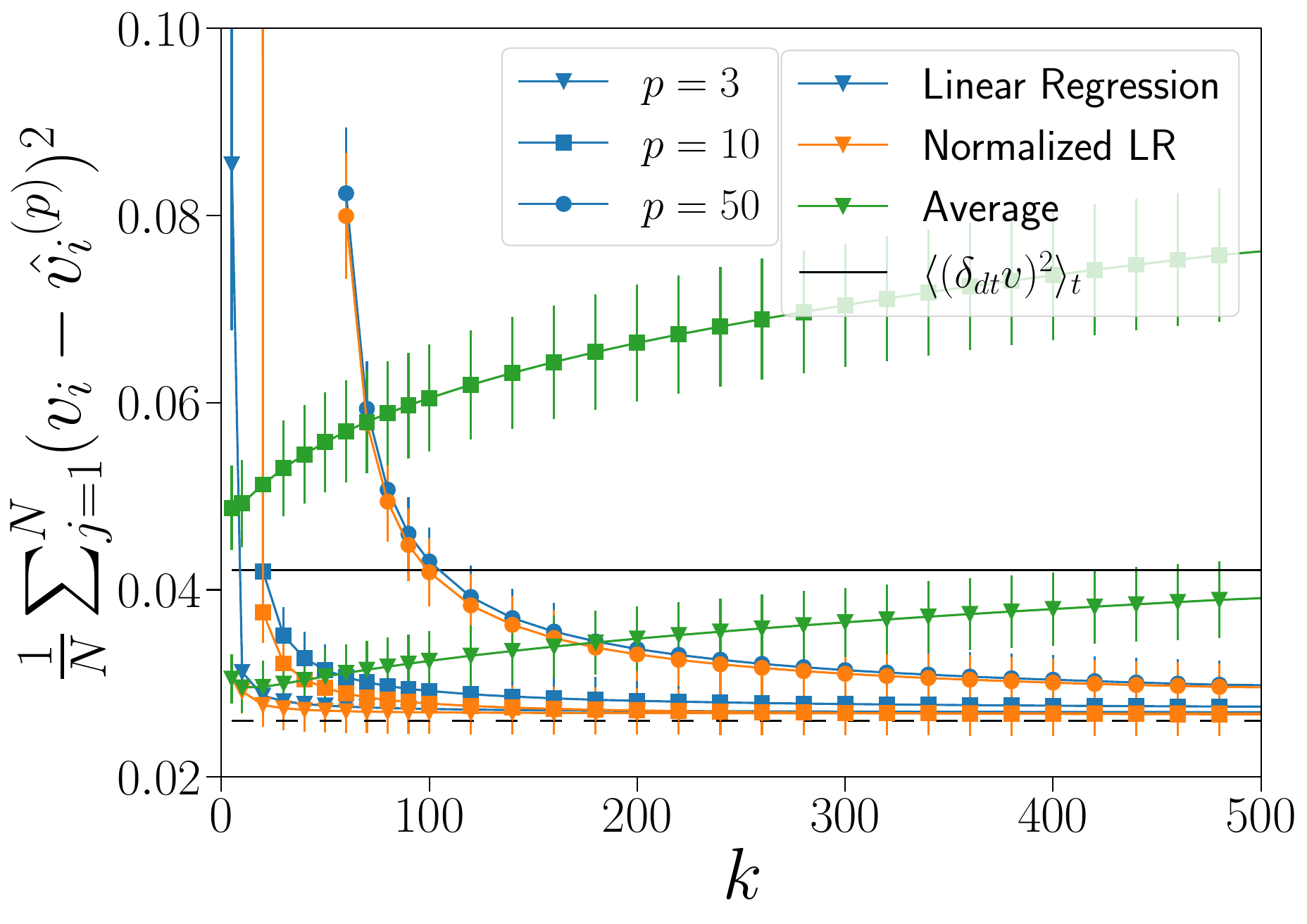}
    \caption{Mean Squared Error $\frac{1}{N}\sum_{j=1}^{N}(v_j - \hat{v_j}^{(p)})^2$ as a function of the number $k$ of neighbors used in the kNN algorithm. The performance is shown for different sizes of the past: $p=3, 10, 50$ and for different methods: Linear Regression (blue), Normalized LR (orange), Averaging LR (green). Vertical lines correspond to the average value and errorbars to the standard deviation computed over $n_{\rm real}$ realizations. Increment variance (continuous black line) is present as a baseline for informative prediction. A dashed line is present to highlight the minimum value towards which the estimator converges. }
    \label{fig:VariancePreds}
\end{figure}

In Figure \ref{fig:VariancePreds} we compare the performance of the three analog prediction methods as a function of $k$ for different values of $p$. 
"Normalized LR" configuration (orange) consistently shows the lowest MSE, indicating superior performance. Regular "LR" (blue) yields similar results. For these two estimators: 1) the MSE increases when $p$ increases due to the curse of dimensionality leading to increased computational complexity and convergence issues and 2) the MSE decreases when $k$ increases until reaching a plateau at $0.026$. "Average" (green) method without linear regression shows higher MSE which increases with the number $k$ of neighbors. 

Increasing the size $p$ of the sub-sequences over $p=3$  does not improve performance, even when $k$ is accordingly increased. This is because three-point statistics already capture most of the dynamics of turbulent flows~\cite{peinkeFokkerPlanckApproach2019b}.

The lower variance in time of innovation ---equal to MSE--- compared to the one of the increment (horizontal continuous black line) indicates that predictions are indeed containing relevant information on the dynamics. The existence of a minimum plateau should be interpreted cautiously since we are using only a single-point velocity measurement, which may not probe the full 3D complexity of turbulence. In line with previous works \cite{lguensatAnalogDataAssimilation2017a}, using linear regression for analogs significantly enhances prediction accuracy, with the linear regression model far outperforming averaging methods.

Based on this performance analysis, we use the "Normalized LR" method in the following, with $p=3$ and $k=70$.

\subsection{Innovation Statistics}

\begin{figure}[t]
    \centering
    \includegraphics[width=0.6\linewidth]{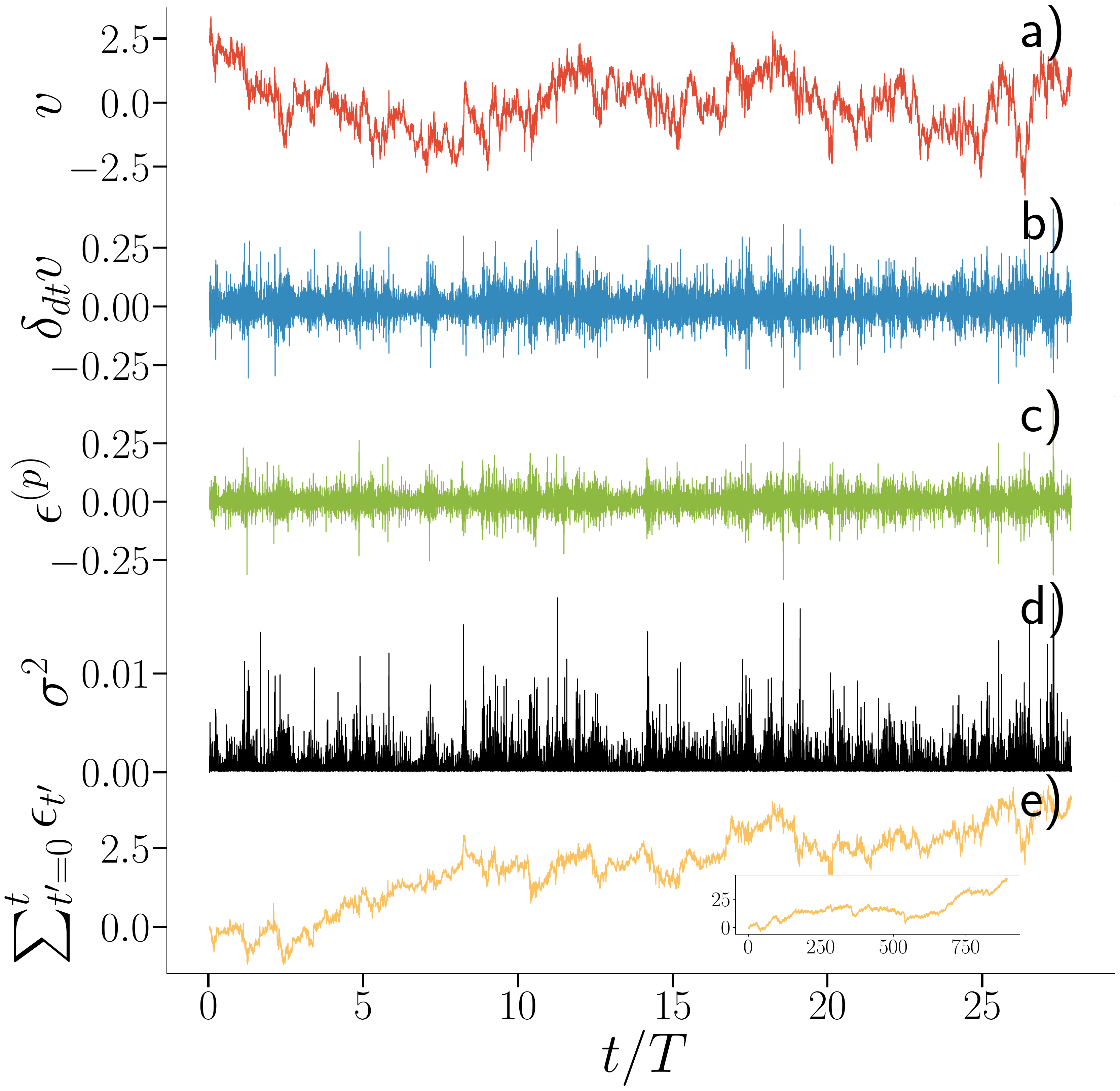}
    \caption{Time series of the Modane velocity $v$ (a), its smallest increment $\delta_{dt} v$ (b), the innovation $\epsilon$ (c), innovation variance $\sigma^2$ (d), and the cumulative sum of innovation $\sum_{t^\prime=0}^{t} \epsilon_{t^\prime}$ (e). The insert shows the cumulative sum of innovation over a longer duration. }
    \label{fig:Signals}
\end{figure}

Figures \ref{fig:Signals}a) and \ref{fig:Signals}b) show respectively the typical evolution of the Modane velocity $v$ and of its  increment $\delta_{dt} v$ over multiple integral scales. The increment exhibits intermittent behavior with calm regions and bursty regions that highlight its non-Gaussianity. Figures \ref{fig:Signals}c) and \ref{fig:Signals}e) show respectively the innovation $\epsilon_t^{(p)}$ and its cumulative sum $\sum_{t'=0}^{t'=t} \epsilon_{t^\prime}^{(p)}$. The innovation mirrors the behavior of the increment but with reduced variance. The concomitant occurence of bursts in both the increment and the innovation time series suggests that extreme events in the increments induced by intermittency lead to an increase in the unpredictability of the velocity values $v_t$ knowing their past $(v_{t-dt},v_{t-2dt},v_{t-3dt})$. This is coherent with previous results illustrating the challenge of predicting extreme events ~\cite{dubrulleKolmogorovCascades2019,buariaExtremeVelocityGradients2019,moffattExtremeEventsTurbulent2021}. Figure \ref{fig:Signals}
d) illustrates the variance of the innovation, highlighting its intermittent nature. 
The variance of the innovation shows spike patterns that match with bursts in the increment and innovations signals.
Furthermore, we observed that $\mathbb{E}[\log(\sigma^2)|\log((\delta_{dt}v)^2))]$ increases with $\log((\delta_{dt}v)^2))$ in a non-linear way. This confirms that regions of high variance of innovation correspond to the presence of intermittent bursts.
This simultaneous peaking underscores that unpredictability is tied with intermittency.
The cumulative sum of the innovation is non-stationary and seems to behave like a random walk, which might indicate that the linear part of the dynamics is correctly predicted in $\hat{v_t}^{(p)}$ and therefore removed from the innovation $\epsilon_t^{(p)}$.

\begin{figure}[t]
    \centering
    \includegraphics[width=0.6\linewidth]{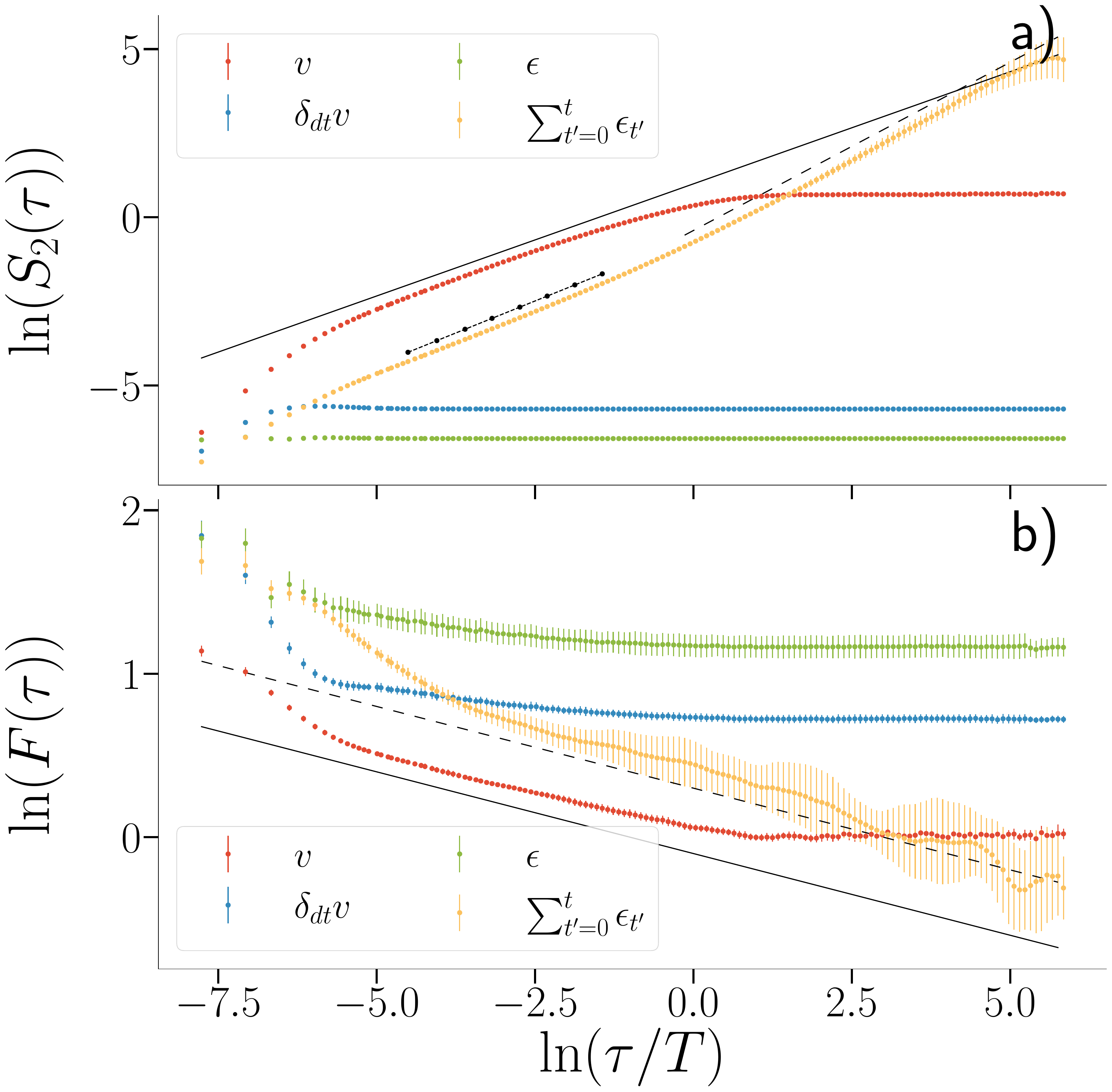}
    \caption{Modane moments: a) second-order structure function $S_2(\bullet)$ and b) flatness $S_4(\bullet)/3\cdot S_2(\bullet)$ for the velocity $v$ (red), the associated smallest increment $\delta_{dt} v$ (blue), innovation $\epsilon$ (green), and innovation sum (yellow). Lines correspond to the average value and errorbars to the standard deviation computed over $n_{\rm real}$ realizations. Computation are done over 10 realizations, each of 892 integral scales. The black lines correspond to power laws: a) continuous:$(\tau/T)^{2/3}$, dashed:$(\tau/T)^{1}$,  round markers :$(\tau/T)^{0.77}$ and b) continuous and dashed:$(\tau/T)^{-0.1}$ with different offsets.}
    \label{fig:Moments}
\end{figure}

Figures \ref{fig:Moments}a) and  \ref{fig:Moments}b) present the evolution of the second-order structure function and  the flatness, with the time-scale $\tau$ of the increment, for the Modane turbulent velocity $v_t$ (red), its increment $\delta_{dt} v$ (blue), the innovation $\epsilon_t^{(p)}$ (green), and its cumulative sum (yellow). 
We relate our results obtained in time to Kolmogorov-Obukhov theories in space
using Taylor's hypothesis~\cite{kolmogorovRefinementPreviousHypotheses1962,obukhovSpecificFeaturesAtmospheric1962}.

The logarithm of the second order structure function $\ln(S_2(\tau/T))$ of the increment $\delta_{dt} v$ and the innovation $\epsilon$ (blue and green curves respectively)
behave very similarly with the time-scale $\tau$. However, it is completely constant across scales for the innovation while it increases at small scales for the increment, until reaching a plateau.

This increase of the second order structure function $S_2$ of the increment at small scales reveals the existence of linear correlations within this process. In contrast, the flat behavior of $S_2$ for the innovation (in green) indicates that linear correlations have been eliminated and accurately predicted by the analog method.

For the velocity (in red), we recognize the three following distinct regions: the integral domain at larger scales, characterized by a plateau and the largest value of $S_2$; the inertial range with the $\tau^{2/3}$ scaling from K41 theory (eq.\ref{eq:Structure}) with $\zeta(2)=2h$ and $h=1/3$) up to intermittent corrections; and the dissipative domain at smaller scales with a $(\tau/T)^{2}$ scaling. For the cumulative sum of the innovation (in yellow), we observe two distinct regions with different power laws: one with an exponent close to ${0.7}$ from the smallest scales up to $\tau\sim T$ and one with an exponent $1$ at larger scales, identical to a Brownian motion.

Finally, the logarithm of the flatness $\ln(F(\tau/T))$ is presented in fig.\ref{fig:Moments}b). The flatness of the increments $\delta_{dt} v$ (in blue) and of the innovation $\epsilon$ (in green) is always larger than one, indicating non-Gaussian behavior at all scales. The flatness of the innovation is larger because simple linear correlations have been predicted, which reduces $S_2$. This implies that the unpredicted extreme events due to intermittency have a greater relative impact, resulting in increased flatness. For the velocity (in red) the flatness tends to 1 at large scales, indicating Gaussian behavior, while being substantially larger at smaller scales, revealing intermittency~\cite{granero-belinchonKullbackLeiblerDivergenceMeasure2018}. The cumulative sum of the innovation (in yellow) shows a power-law behavior for scales larger than $\ln(\tau/T) \sim -3$ with an exponant close to $-0.1$. For smaller scales, the evolution of the flatness is faster. Such behavior indicates the persistence of higher-order correlations within the innovation signal, despite the removal of most, if not all, linear correlations.

\section{Conclusion}

In this study, we used analog-based predictions of turbulent velocity measurements from the Modane wind tunnel and computed the innovation in order to analyze the predictability of turbulent velocity. A first application confirmed that linear regression-based analog prediction better accounts for complex dynamics such as turbulence, compared to the standard averaging method, as suggested by state-of-the-art studies. Our analysis then illustrated that extreme events in the turbulent velocity gradient correspond to extreme events in the innovation, indicating a direct relationship between intermittency-related bursts of the velocity signal and harder to predict velocity values. These extreme events coincide also with high values of the uncertainty of the prediction. The correspondence between high values of the innovation and high values of the uncertainty of the prediction could be illustrative of the intrinsic stochasticity of turbulence. Thus in turbulence, predictions should better exploit the information of the pdf $q(v_{t}|v_{t-p dt:t-dt})$, not considering only its first order moment.

Moreover, a multi-scale analysis revealed that the analog-based estimator of turbulent velocity effectively predicts and removes most linear dependencies in the turbulent velocity signal, as evidenced by the flat second-order structure function of the corresponding innovation. The second-order structure function of the cumulative sum of the innovation does not exhibit the typical dissipative and integral domains. Instead, it shows two distinct regions with different power laws: one with a slope close to 0.7 and another with a slope 1. This suggests that while the estimator captures linear dynamics, it struggles with high-order dependencies. The flatness analysis further confirms that high-order dependencies remain largely intact, as indicated by the same power law of exponent -0.1 for the cumulative sum of the innovation as for the turbulent velocity itself. These findings highlight the limitations of the current analog-based estimator in fully capturing high-order interactions and extreme events. This is an illustration of the limitations of the hypotheses behind analog-based forecasting when applied to turbulence. Future research should focus on introducing more physical, e.g. multifractal, models in the selection of analogs to better predict high-order interactions and better capture the full complexity of turbulent flows.


\acknowledgments

The code used for the estimation of analog-based predictions and innovations is available in :
https://gitlab.imt-atlantique.fr/e22froge/multi-scale-causality \\

The authors wish to thank P. Tandeo for enriching discussions. This work was supported by the French National Research Agency (ANR-21-CE46-0011-01), within the program ``Appel \`a projets g\'en\'erique 2021''.

\bibliographystyle{elsarticle-num} 
\bibliography{ArticleInnovation}

\end{document}